\documentclass[a4paper,11pt]{article}
\usepackage{pos}

\title{
Refactorization of endpoint divergencies \\for the ${\cal O}_7$ contribution  to $\bar B_s \to \mu^+\mu^-$ 
}
\ShortTitle{Refactorization of endpoint divergencies for the ${\cal O}_7$ contribution  to $\bar B_s \to \mu^+\mu^-$}

\author*[]{Nicolas Seitz}

\affiliation[]{Theoretische Physik 1, Center for Particle Physics Siegen (CPPS), Universit\"at Siegen,\\
Walter-Flex-Stra{\ss}e 3, D-57068 Siegen, Germany}

\emailAdd{nicolas.seitz@uni-siegen.de}

\abstract{
We report on the construction of a factorization theorem for the contribution of the electromagnetic dipole operator ${\cal O}_7$ to the $\bar B_s \to \mu^+\mu^-$ decay amplitude. The leading-order contribution from a QED box diagram features a double-logarithmic enhancement associated to the different rapidities of the light quark in the $\bar B_s$-meson and the energetic muons in the final state. 
We analyse the cancellation of the related endpoint divergences appearing in individual momentum regions, and show how the rapidity logarithms can be isolated by suitable subtractions applied to the corresponding bare factorization theorem.
This allows us to include in a straightforward manner the QCD corrections arising from the renormalization-group running of the hard matching coefficient, the hard-collinear scattering kernel, and the $\bar B_s$-meson distribution amplitude.
}

\FullConference{%
  21st Conference on Flavor Physics and CP Violation (FPCP 2023) \\
  29 May - 2 June 2023\\
  IP2I - Lyon University, Lyon, France\\[1em]
  Preprint numbers: SI-HEP-2023-16, P3H-23-047
}

\begin{document}

\maketitle

\section{Introduction}

The rare $\bar B_s \to \mu^+\mu^-$ decay is of strong interest from a phenomenological point of view. Since the decay amplitude at leading-order stems only from the operator ${\cal O}_{10}$ it is possible to constrain on the new physics contributions to the ${\cal C}_{10}$ Wilson coefficient. The decay rate can be measured precisely due to its leptonic nature. Recent results are published in Ref.~\cite{ATLAS:2018cur, CMS:2022mgd, LHCb:2021awg}. On the theory side the transition $b \to s\ell^+\ell^-$ is a flavor-changing neutral current and hence loop-induced. Very precise predictions have been calculated, including higher-order QCD and electroweak corrections~\cite{Buras:2012ru, Bobeth:2013uxa, Buras:2013uqa, DeBruyn:2012wk}. In Ref.~\cite{Beneke:2017vpq, Beneke:2019slt} at the level of intended precision also non local QED corrections have been calculated. Comparing theoretical and experimental results is therefore a very precise test of the Standard Model (SM) of particle physics and allows also for the search of physics beyond the SM. In theses characteristics the process $\bar B_s \to \mu^+\mu^-$ is complementary to other decay modes like $\bar B \to X_s \gamma$ and $B \to K^{(*)}\ell\ell$. This letter is based on our article in Ref.~\cite{Feldmann:2022ixt}.

\section{The \texorpdfstring{$\bar B_s \to \mu^+\mu^-$} decay amplitude}

Starting from the SM and integrating out heavy bosons $W$, $Z$, $H$ and the heavy top-quark, one arrives at the so-called {\emph{effective weak Hamiltonian}}. The most important operators for the $\bar B_s \to \mu^+\mu^-$ process are \cite{Chetyrkin:1996vx},
\begin{eqnarray*}
\qquad \qquad  {\cal O}_{10} &=& (\bar s \gamma_\mu P_L b)(\bar \mu \gamma^\mu\gamma_5 \mu) \,,\\
\qquad \qquad  {\cal O}_{9} &=& (\bar s \gamma_\mu P_L b)(\bar \mu \gamma^\mu \mu) \,,\\[0.3em]
\qquad \qquad {\cal O}_7 &=& \frac{e}{16\pi^2} m_b (\bar{s}_L \sigma_{\mu \nu} b_R) F^{\mu \nu} \,.
\end{eqnarray*}
At leading order only the operator ${\cal O}_{10}$ contributes. At this level the hadronic uncertainty to the decay amplitude stems from the decay constant $f_{B_s}$. Furthermore, one finds a helicity suppression, which is evident in the decay amplitude with a linear muon mass $m$. To calculate QED corrections at one-loop order one has to take into account photon exchange between the light particles of the process. Together with the ${\cal O}_{10}$ operator of the tree-level diagram, the ${\cal O}_{9}$ and ${\cal O}_7$ operators then do contribute at this order \cite{Beneke:2017vpq, Beneke:2019slt}. The helicity suppression becomes lifted by the strange-quark propagator and the decay amplitude contains a single-logarithmic enhancement from the ${\cal O}_{9}$ and a double-logarithmic enhancement from the ${\cal O}_7$ operator. Thus, QED effects can become phenomenologically relevant, but nethertheless remain small due to a supression in $\alpha$. In this work we focus on the electric dipole operator ${\cal O}_7$. The effects coming from this operator are particulary interesting to study. The ${\cal O}_7$ contribution, which is diagrammatically shown in Fig.~\ref{fig:box}, is also of importance from an conceptual point of view. By the exchange of the photon the decay amplitude becomes sensitive to the momentum distribution of the light quark inside the $\bar B_s$-meson. Calculating the diagram in Fig.~\ref{fig:box} with the method of regions \cite{Beneke:1997zp,Smirnov:1998vk} leads to endpoint divergent convolution integrals. These endpoint divergences have to be regularised by rapidity regulators. We studied intensively the cancellation of the endpoint poles between the different momentum regions depending on the choice of the regulator. One of the endpoint configurations of the muon propagator results in the already mentioned double-logarithmic enhancement. Endpoint divergent convolution integrals make the formulation of a factorization theorem non-trivial. In our work we provided a factorization theorem for the ${\cal O}_7$ contribution that is free of endpoint divergences and valid to all orders in QCD. In the context of QCD factorization the topic of endpoint-logarithms is a very much discussed topic in recent times, see e.g.\ \cite{Boer:2018mgl,Liu:2018czl,Liu:2019oav,Liu:2020wbn,Beneke:2022obx,Bell:2022ott,Cornella:2022ubo,Hurth:2023paz}. We were able to calculate the leading-logarithmic QCD corrections using renormalization group improved pertubation theory.

\section{\texorpdfstring{${\cal O}_7$} contribution to the decay amplitude}
\begin{figure}[t] 
\begin{center}
	\includegraphics[width=0.60\textwidth]{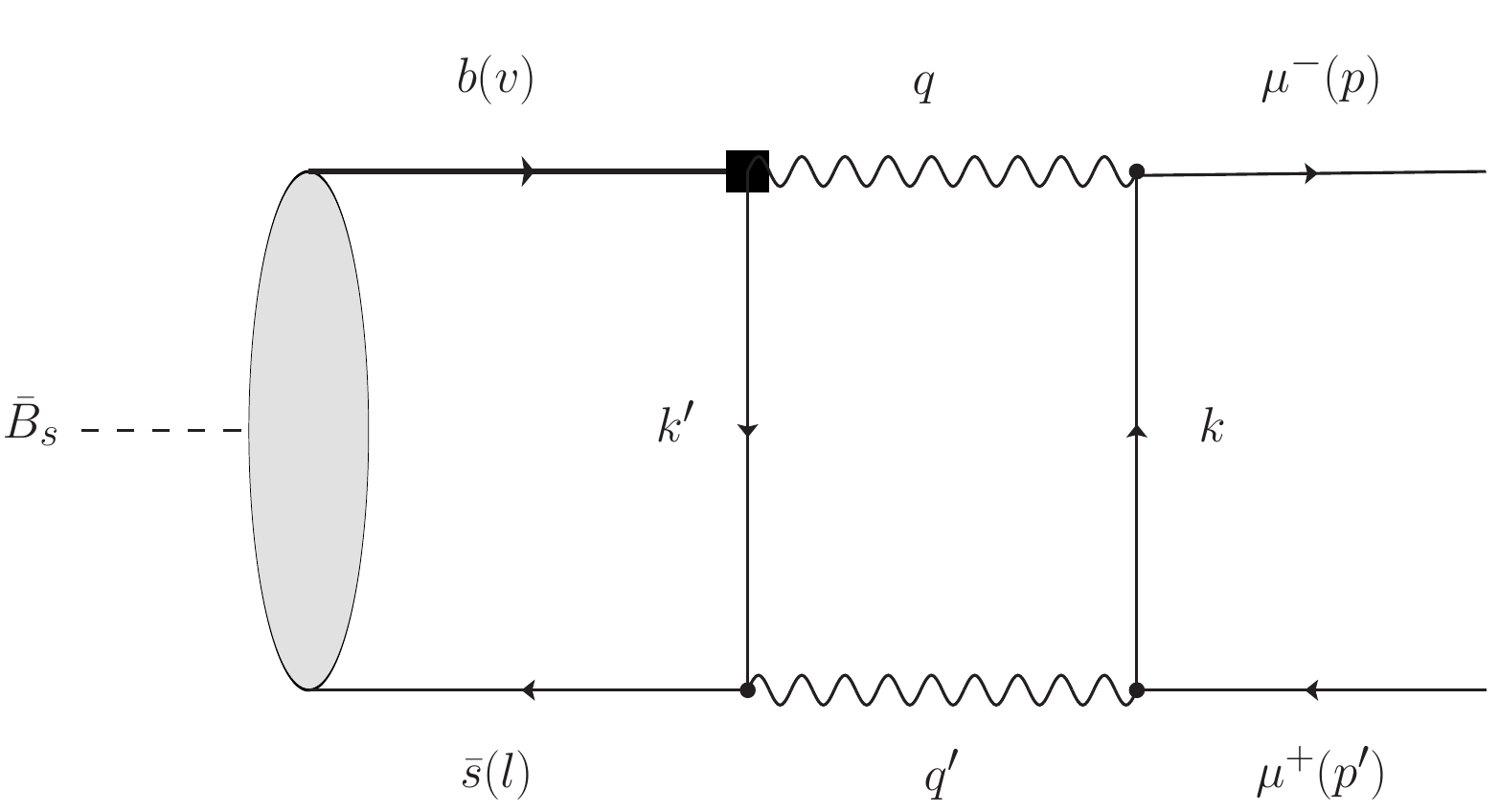}
\end{center}
\caption{\label{fig:box} Leading order QED box diagram that contributes to the $\bar B_s \to \mu^+\mu^-$ decay amplitude from the operator ${\cal O}_7$. There exists a second diagram with the crossed photons, that is not shown in this figure.
}
\end{figure}
The QED correction at ${\cal O}(\alpha)$ to the $\bar B_s \to \mu^+\mu^-$ decay amplitude has already been calculated and can be found in Ref.~\cite{Beneke:2017vpq}. For our work the relevant part of the decay amplitude, that stems from the operator ${\cal O}_7$ can be written in the form
\begin{eqnarray}
    i {\cal M}(\bar B_s \to \mu^+\mu^-) \Big|_{{\cal O}_7}^{\rm LO} &=&
    - \frac{\alpha}{2\pi} \, Q_\ell^2 \, Q_s \, C_7^{\rm eff} \, m \, M  f_{B_s} \, {\cal N} \, 
    \left[\bar u(p) \, (1+\gamma_5) \, v(p') \right] {\cal F}^{\rm LO}(E,m)
    \,,
\end{eqnarray}
with the constant
\begin{eqnarray}
    {\cal N} &=& V_{tb} \, V_{ts}^* \, \frac{4G_F}{\sqrt2} \, \frac{\alpha}{4\pi} \,.
\end{eqnarray}
The corresponding diagram is shown in Fig.~\ref{fig:box}, where all hadronic information is included in the form factor
\begin{eqnarray}
    {\cal F}^{\rm LO} (E,m) &=& \int_0^\infty \frac{d\omega}{\omega} \, \phi_+(\omega) 
    \left[ \frac12 \, \ln^2 \frac{m^2}{2E\omega} +  \ln \frac{m^2}{2E\omega} + \frac{\pi^2}{3} \right] \,.
    \label{eq:LO} 
\end{eqnarray}
This form factor consists of the so-called light cone distribution amplitude (LCDA) of the $\bar B_s$-meson $\phi_+(\omega)$, which shows that through the QED correction the process is sensitive to the momentum distribution of the spectator quark in the $\bar B_s$-meson. The argument of the LCDA is the light-cone projection of the strange-quark momentum $l$ (compare to Fig.~\ref{fig:box}) which has the power counting $\langle \omega \rangle \sim {\cal O}(\Lambda_{\rm had})$.
\section{Momentum regions in the 1-Loop QED diagram}
In this section we calculate the diagram shown in Fig.~\ref{fig:box} using the method of regions approach. For this purpose we introduce light-cone coordinates for the muon momentum, 
    $$
        k^\mu = (\overline{n} \cdot k) \frac{n^\mu}{2} + (n \cdot k) \frac{\overline{n}^\mu}{2} + k_\perp^\mu
    $$
with $n^2=\bar n^2=0$ and $n\cdot \bar n=2$. In this process there is a hard scale $(\bar n \cdot p) = (n\cdot p')\simeq 2E = M$, where $E$ is the energy of the muons in the $\bar B_s$ rest frame and $M$ the meson mass. The soft scale in this process is the intrinsic QCD scale $\Lambda_{\rm QCD}$. The projection of the strange-quark momentum  $\omega=(\bar n\cdot \ell)$ and the muon mass $m$ are counted to be of this power. In the spirit of the method of regions we define a small expansion parameter $\lambda$ by
    \begin{equation}
        \lambda^2 = \frac{m^2}{2E\omega} \sim {\cal O}\left( \frac{\Lambda_{\rm QCD}}{m_b}\right) \sim {\cal O}\left( \frac{m_\mu}{m_b} \right).
    \end{equation}
    
\begin{table}[t]
\begin{center}
    \begin{tabular}{c||c|c|c|c||c}
    \hline 
    region & muon  & upper photon & $\bar s$-quark & lower photon & regulator 
    \\
    \hline \hline 
 $\overline{hc}$  
 & $k \sim (\lambda^2,\lambda,1)$ 
 & $q \sim (1,\lambda,1)$
 & $k' \sim (\lambda^2,\lambda,1)$
 & $q' \sim (\lambda^2,\lambda,1)$
 & ${\cal R}_{a,b}$
 \\
 \hline 
 $\overline{c}$  
 & $k \sim (\lambda^4,\lambda^2,1)$ 
 & $q \sim (1,\lambda^2,1)$
 & $k' \sim (\lambda^2,\lambda^2,1)$
 & $q' \sim (\lambda^4,\lambda^2,1)$
 & ${\cal R}_{a,b}$
 \\
 \hline 
 $s$  
 & $k \sim (\lambda^2,\lambda^2,\lambda^2)$ 
 & $q \sim (1,\lambda^2,\lambda^2)$
 & $k' \sim (\lambda^2,\lambda^2,1)$
 & $q' \sim (\lambda^2,\lambda^2,1)$
 & ${\cal R}_{a,b}$
 \\
 \hline 
 $\overline{sc}$  
 & $k \sim (\lambda^3,\lambda^2,\lambda)$ 
 & $q \sim (1,\lambda^2,\lambda)$
 & $k' \sim (\lambda^2,\lambda^2,1)$
 & $q' \sim (\lambda^3,\lambda^2,1)$
 & ${\cal R}_{b}$
 \\
 \hline 
 \hline 
    \end{tabular}
\end{center}
\caption{\label{tab:regions} In this table we show the relevant impulse regions for all propagators that contribute to the diagram in Fig.~\ref{fig:box}. The anti-soft-collinear region only contributes if the regulator (b) is chosen. The power-counting of the three different momentum components in the light-cone notation takes the form $\left\{ ( \bar n \cdot k) , \ k_\perp , (n \cdot k) \right\} \sim (\lambda^a,\lambda^b,\lambda^c) \, m_b$. From this we derive the short-hand notation $\; k^\mu \sim  (\lambda^a,\lambda^b,\lambda^c) \;$ used in this table and throughout the letter.
}
\end{table}
The different regions that have to be considered are summarized in Table \ref{tab:regions}. By expanding the loop integral into the different regions additional divergences - the so-called endpoint divergences - show up, which must be regularized. To do so, one can choose different options. Two possible options are:
\begin{eqnarray}
	\mbox{option (a)}  &:& \qquad 
	{{\cal R}_a(k)} = \left(\frac{\nu^2}{-(n \cdot k)(\bar n\cdot l) + i 0} \right)^\delta \,, 
	\\
	\mbox{option (b)} &:& \qquad 	{{\cal R}_b(k)} =	\left( \frac{\nu^2}{(\bar n \cdot k)(n \cdot p') - (n\cdot k) (\bar n \cdot l) + i 0}\right)^\delta \, .
\end{eqnarray}
Note, that depending on the choice different sets of regions have to be calculated. If one decides on option (a) the regions anti-hard-collinear $\overline{hc}$, anti-collinear $\overline{c}$ and soft $s$ contribute. With option (b) additionally the region anti-soft-collinear $\overline{sc}$ has to be calculated to obtain the correct result for the diagram in Eq.~(\ref{eq:LO}).

From Table~\ref{tab:regions} we find, that the virtuality is in all momentum regions $(k^\prime)^2 \sim \lambda^2$. Because of this, we can already simplify the strange-quark propagator (third propagator in the following equation) and can write the loop-integral in the form
\begin{equation}
\begin{aligned}
	I(\omega) &= \int \! d (\bar n \cdot k) \, \int \! d (n \cdot k) \, 
	\int \widetilde{dk_\perp} \,  \cdot \frac{1}{(p-k)^2+i0}
        \cdot\frac{1}{(p'+k)^2+i0} 
	\cr &\times
	 \frac{(2E + n \cdot k) \, 2E\omega}{(2E+n \cdot k)(\bar n \cdot k-\omega) + k_\perp^2+i0}
	\cdot \frac{2E + n \cdot k}{(n \cdot k)(\bar n \cdot k) + k_\perp^2 -m^2 +i0} 
	\, \Bigg|_{\mbox{\scriptsize leading power}} \,.
\label{Iomega}
\end{aligned}
\end{equation}
This integral will simplify further by expanding it to leading power for the different momentum regions. In the following we choose rapidity regulator (a). 
\subsection{The anti-hard-collinear region}
The anti-hard-collinear region is given by the momentum scaling $k  \sim  (\lambda^2,\lambda,1)$. Expanding the integral in Eq.~(\ref{Iomega}) in this region we find at leading power in $\lambda$:
\begin{equation}
\begin{aligned}
	I_{\overline{hc}}(\omega) &= \int d(\bar n \cdot k) \, \int d (n \cdot k) \, 
	\int \widetilde{dk_\perp} \, \\[0.3em]
	&\times \, \frac{1}{- 2E \, (n \cdot k)+i0}
	\cdot\frac{1}{(2E+n \cdot k)(\bar n \cdot k) + k_\perp^2+i0} \\[0.3em]
	&\times \, 
	\frac{(2E + n \cdot k) \, 2E\omega}{(2E+n \cdot k)( \bar n \cdot k- \omega) + k_\perp^2+i0}
	\cdot \frac{2E + n \cdot k}{(n \cdot k)(\bar n \cdot k) + k_\perp^2 +i0} \, .
\end{aligned}
\end{equation}
Performing first the $(\bar n \cdot k) $ integration by using the residue theorem and afterwards the $k_\perp$ integral, we find with $(n \cdot k) = - 2 \, E \, u$ for the convolution integral,
\begin{eqnarray}
	I_{\overline{hc}}(\omega) &=&
    \int_0^1 \frac{du}{u} \,  H_1^{(0)}(u) \,
	\bar J_1^{(1)}(u;\omega)
	\,,
\end{eqnarray}
where $
	H_1^{(0)}(u) = 1 
$
is the leading-order value of the hard SCET-function (hard matching coefficient for the 
$b \to s\gamma$ tensor current with an energy transfer $(1-u) E$) and
\begin{eqnarray}
	 \bar J_1^{(1)}(u;\omega) 
	 = -  \Gamma(\epsilon)  \left( \frac{\mu^2 e^{\gamma_E}}{2E\omega u(1-u)} \right)^\epsilon (1-u)
\label{H1J1}
\end{eqnarray}
is the jet-function at leading-order. The endpoint divergence in this region stems from the eikonal photon propagator und shows up as the $1/u$ term in the convolution integral. Note, that the endpoint divergence is already regularized by the dimensional regulator $\epsilon$ through the $u^{-\epsilon}$ term in the jet function. For this reason no additional rapidity regulator is needed and we can set $\delta$ to zero. Integrating over the remaining momentum fraction $u$ and expanding in $\epsilon$ one finds
\begin{eqnarray}
	I_{\overline{hc}}(\omega) = 
		\frac{1}{\epsilon^2} + \frac1{\epsilon} \, \ln \frac{\mu^2}{2E\omega} + \frac12 \, \ln^2\frac{\mu^2}{2E\omega} -  \frac{\pi^2}{12}
		+ \frac{1}{\epsilon} + \ln \frac{\mu^2}{2E\omega} + 2 \, .
\end{eqnarray}
\subsection{The anti-collinear region}
The anti-collinear region is given by the momentum scaling $k  \sim  (\lambda^4,\lambda^2,1)$. Expanding the integral in Eq.~(\ref{Iomega}) in this region we find at leading power in $\lambda$:
\begin{equation}
\begin{aligned}
	I_{\bar c}(\omega) &= \int \! d (\bar n\cdot k) \, \int \! d (n \cdot k) \, 
	\int \widetilde{dk_\perp} \,  \\[0.3em]
	&\times \, \frac{1}{-2E\, (n \cdot k)+i0}
	\cdot\frac{1}{(2E+n \cdot k)(\bar n \cdot p' +\bar n \cdot k)+ k_\perp^2+i0}  \\[0.3em] 
	&\times \,
	\frac{2E\omega}{-\omega +i0}
	\cdot \frac{2E + n \cdot k}{(n \cdot k)(\bar n \cdot k) + k_\perp^2 - m^2 +i0} \left( \frac{\nu^2}{- \omega \, (n \cdot k) + i0}\right)^\delta \, .
\end{aligned}
\end{equation}
Again we perform first the $(\bar n \cdot k) $ and afterwards the $k_\perp$ integral. With $(n \cdot k) = - 2 \, E \, u$ the result can be written as
\begin{eqnarray}
	I_{\overline{c}}(\omega) &=&
    \bar J_{2}^{(0)}(1,\omega) \, \int_0^1 \frac{du}{u} \,  H_1^{(0)}(u) \,
	\bar C^{(1)}(u;\omega) \,,
\end{eqnarray}
where we defined the leading-order term in the jet function 
for the anti-hard-collinear strange-quark propagator with $(\bar n\cdot k') = z \, \omega$, 
\begin{eqnarray} 
\bar J_2^{(0)}(z,\omega) &=& \frac{1}{z} \label{barJ2} \,,
\end{eqnarray}
such that the overall factor $1/\omega$ --~that appears in the convolution with the $\bar B_s$-meson LCDA~--
has been factored out.
At leading order the collinear function is defined as
\begin{eqnarray}
   \bar C^{(1)}(u;\omega)  
    &=& 
	 \Gamma(\epsilon) \,
\left(\frac{\mu^2 e^{\gamma_E}}{m^2}\right)^{\epsilon}
		\left( \frac{\nu^2}{2E\omega} \right)^\delta
	\left( 1-u \right)^{1-2\epsilon} \, u^{-\delta} \,.
	\label{C1}
\end{eqnarray}
In the convolution integral again an endpoint divergence $1/u$ from the eikonal photon propagator shows up. The endpoint divergence is not regulated without the rapidity regulator $\delta$, so we need to keep the term $u^{-\delta}$ in the collinear function. Integrating the last momentum fraction integral $u$ and then expanding the result first in $\delta$ and afterwards in $\epsilon$ we find
\begin{eqnarray}
	I_{\overline{c}}(\omega) &=& 
	\left( -\frac{1}{\delta} - \ln \frac{\nu^2}{2E\omega} \right) 
	\left( \frac{1}{\epsilon} + \ln \frac{\mu^2}{m^2} \right)
	+ \frac{\pi^2}{3} 
	- \frac{1}{\epsilon} - \ln \frac{\mu^2}{m^2} -2 \, .
\end{eqnarray}
\subsection{The soft region}
The soft region is given by the momentum scaling $k  \sim  (\lambda^2,\lambda^2, \lambda^2)$. Expanding the integral in Eq.~(\ref{Iomega}) in this region we find at leading power in $\lambda$:
\begin{equation}
\begin{aligned}
	I_s(\omega) &= \int d (\bar n \cdot k)
\, \int d (n \cdot k) \, 
\int \widetilde{dk_\perp} \, 
 \cdot \frac{1}{-2E \, (n \cdot k)  +i0}
\cdot\frac{1}{2E \, (\bar n \cdot k)+i0}
\cr 
&\times
\frac{2E \omega}{\bar n \cdot k- \omega + i0}
\cdot \frac{2E}{(n \cdot k)(\bar n \cdot k) + k_\perp^2 -m^2 +i0} \,
{\cal R}_{a}(k)
\,,
\end{aligned}
\end{equation}
In this region we can perform first the $k_\perp$ integration. Afterwards we have to examine the analytic structure in the $(\bar n \cdot k)$-plane and consider the poles and cuts respectively. After this procedere, one ends up with
\begin{equation}
\tilde I_s(\omega) = 
 H_1^{(0)}(0) \, \int_0^\infty \frac{du}{u} \, \int_0^\infty \frac{d\rho}{\rho} \, S^{(1)}(u,\rho;\omega) \, \bar J_2^{(0)}(1+\rho,\omega) \, .
\end{equation}
Here we find with $(n \cdot k) = - 2 \, E \, u$ and $(\bar n \cdot k) = - \omega \, \rho$ the same jet function as already defined in the anti-collinear region and a soft function
\begin{equation}
    S^{(1)}(u,\rho;\omega) = \theta(u\rho -\lambda^2) \left( \frac{\mu^2  e^{\gamma_E}}{2E\omega} \right)^\epsilon \left( \frac{\nu^2}{2uE\omega} \right)^\delta   \frac{(u\rho -\lambda^2)^{-\epsilon}}{ \Gamma(1-\epsilon)} \, \,,
    \label{S1}
\end{equation}
originating from the discontinuity of the muon propagator.
The convolution integral is again endpoint divergent, which stems from small values $(n \cdot k)$ and $(\bar n \cdot k)$ in the two eikonal photon propagators. This shows up as the $1/u$ and $1/\rho$ terms in the convolution integrals and again leads to the fact, that we have to keep the rapidity regulator $\delta$. After calculating the remaining longitudinal integrals and afterwards expanding first in $\delta$ and subsequently in the dimensional regulator $\epsilon$, we find 
\begin{eqnarray}
\tilde I_s(\omega) &=&
\left( \frac{1}{\delta} + \ln \frac{\nu^2}{m^2}\right)\left( \frac{1}{\epsilon} + \ln \frac{\mu^2}{m^2}\right) - \frac{1}{\epsilon^2} - \frac{1}{\epsilon} \, \ln \frac{\mu^2}{m^2}  - \frac12 \, \ln^2 \frac{\mu^2}{m^2} + \frac{\pi^2}{12}
\,.
\end{eqnarray}
\subsection{Total result for the 1-Loop QED diagram}
As already stated, the total result of the diagram in Fig.~\ref{fig:box} together with rapidity regulator (a) is given by the sum of the anti-hard-collinear, the anti-collinear, and the soft region, which leads to the net result
\begin{eqnarray}
I(\omega) &=& I_{\overline{hc}}(\omega) + I_{\overline c}(\omega) + \tilde I_s(\omega) 
	= \frac12 \, \ln^2 \frac{m^2}{2E\omega} 
	+ \ln \frac{m^2}{2E\omega} + \frac{\pi^2}{3} \,.
\end{eqnarray}
All poles in $\epsilon$ and $\delta$ cancel out and we reproduce the result shown in the form factor in Eq.~(\ref{eq:LO}). Note the double-logarithmic enhancement and the single-logarithmic term. The single-logarithmic enhancement stems from the cancellation of the $1/\epsilon$ poles between the anti-collinear and the anti-hard-collinear regions. 
\section{The bare QCD factorization theorem}
The starting point to render the convolution integrals endpoint finite, is the so-called bare QCD factorization theorem. The form factor ${\cal F}(E,m)$ is the sum of the convolution integral of the anti-hard-collinear $\overline{hc}$, the anti-collinear $\overline{c}$ and the soft $s$ region. This leads to the expression
\begin{eqnarray}
{\cal F}(E,m) &=& \int_0^\infty \frac{d\omega}{\omega} \, \phi_+(\omega) 
\Bigg\{ \int_0^1 \frac{du}{u} \, H_1(u) \, \bar J_1(u;\omega) \,
\cr 
&& \qquad {} + \bar J_2(1,\omega) \, \int_0^1 \frac{du}{u} \, H_1(u) \, \bar C(u;\omega) 
\cr 
&&
\qquad {} + H_1(0) \, \int_0^\infty \frac{du}{u} \, \int_0^\infty \frac{d\rho}{\rho} \, S(u,\rho;\omega) \, \bar J_2(1+\rho,\omega)  \Bigg\}_{\rm bare} \, .
\label{eq:bare}
\end{eqnarray}
Every line in this QCD factorization theorem contains endpoint divergent convolutions, which are regulated by finite $\epsilon$ and $\delta$. To include QCD corrections to the process on the quark side of the diagram in Fig.~\ref{fig:box} we write the SCET functions as
\begin{align}
H_1(u) & = H_1^{(0)}(u) + {\cal O}(\alpha_s) \, , \\
\bar J_1(u;\omega) & = \bar J_1^{(1)}(u;\omega) + {\cal O}(\alpha_s) \, , \\
\bar J_2(z) & = \bar J_2^{(0)}(z) + {\cal O}(\alpha_s) \,,
\end{align}
with $H_1^{(0)}(u)=1$, and $\bar J_1^{(1)}(u;\omega)$, $\bar J_2^{(0)}(z)$ given in 
Eqs.~(\ref{H1J1}) and~(\ref{barJ2}), respectively. The collinear function $\bar C(u;\omega) $ and the soft function $S(u,\rho;\omega)$ cannot include any QCD corrections due to the fact, that they stem from photon propagators and the muon propagator. Not included into this factorization theorem are QED corrections from the subprocess $\gamma^*\gamma^* \to \mu^+\mu^-$, so we set the functions $\bar C(u;\omega) $ and $S(u,\rho;\omega)$ to their leading order expressions.
\section{Construction of the renormalized QCD factorization theorem}
Since the total result is, as we have seen above, free of any endpoint divergences, we aim to rewrite the factorization in such a way, that all lines are free of endpoint divergences. After this procedure we can get rid of the rapidity regulator before the integrations and set $\delta = 0$.
This can be achieved by the so-called refactorization procedure. To perform this, we need as an essential ingredient the refactorization relation Ref.~\cite{Boer:2018mgl,Liu:2019oav,Bell:2022ott}
\begin{eqnarray}
&& \left[\left[\bar C(u;\omega)\right]\right] \equiv \bar C(u;\omega)\big|_{u\to 0}
= \int_0^\infty \frac{d\rho}{\rho} \, S(u,\rho;\omega) + {\cal O}(\alpha) \, , 
\label{refac}
\end{eqnarray}
and additionally use the fact, that scaleless integrals vanish in dimensional regularisation
\begin{eqnarray}
\int_0^\infty \frac{du}{u} \, \int_0^\infty \frac{d\rho}{\rho} \, S(u,\rho;\omega) &=& 0 \,.
\end{eqnarray}
Note, that the refactorization condition is valid for all orders in $\alpha_s$, but is only valid for the leading order in $\alpha$. Applying these two formulas to the bare factorization theorem we gain the renormalized factorization theorem
\begin{eqnarray}
{\cal F}(E,m) &=& \int_0^\infty \frac{d\omega}{\omega} \, \phi_+(\omega) 
\Bigg\{ \int_0^\infty \frac{du}{u} \Big[ H_1(u) \, \bar J_1(u;\omega) \, \theta(1-u) 
\cr && \qquad \qquad \qquad \qquad   {}
- H_1(0) \, \bar J_2(1,\omega) \, \theta(u-1) \, \int_1^\infty\frac{d\rho}{\rho} \, S(u\rho;\omega) \Big]_{\lambda^2 \to 0} 
\cr 
&& \qquad {} + \bar J_2(1,\omega) \, \int_0^1 \frac{du}{u} \Big[ H_1(u) \,  \bar C(u) - H_1(0) \left[\left[ \bar C(u)\right]\right] \Big]
\nonumber\\[0.5em]
&& \qquad {} +H_1(0) \, \int_0^\infty \frac{du}{u} \, \int_0^\infty \frac{d\rho}{\rho} \left[ \bar J_2(1+\rho,\omega) -  
\theta(1-\rho) \, \bar J_2(1,\omega) \right] S(u \rho;\omega)
\nonumber\\[0.3em]
&&  \qquad {} + \bar J_2(1,\omega) \, H_1(0) \, \int_0^1 \frac{du}{u}\, \int_0^1 \frac{d\rho}{\rho} \,  S(u \rho;\omega) \Big|_{\rm \lambda^2\to 0} \Bigg\} \,.
\label{fact}
\end{eqnarray}
We recognize, that the factorization theorem now consists of four lines. All of these lines are free of endpoint divergences, so the rapidity regulator $\delta$ can be dropped. The last line is new compared to the bare factorization theorem and contains the double-logarithmic enhancement as
\begin{eqnarray}
	{\cal F}^{\rm LO}(E,m)\Big|_{\rm double-log} &=& 
	\int \frac{d\omega}{\omega} \, \phi_+(\omega) \, \int\limits_0^1 \frac{du}{u} \, 
	\int\limits_0^1 \frac{dv}{v} \, 
	\theta(2 E  \omega  \, uv-m^2)
	\,.
\end{eqnarray}
This can also be obtained by calculating soft-collinear region $\overline{sc}$ as shown in Table~\ref{tab:regions} with additional longitudinal cut-offs \cite{Liu:2018czl}.
\section{Leading-Logarithmic QCD Corrections}
\begin{figure}[t]

\begin{center} 
\includegraphics[width=0.49\textwidth]{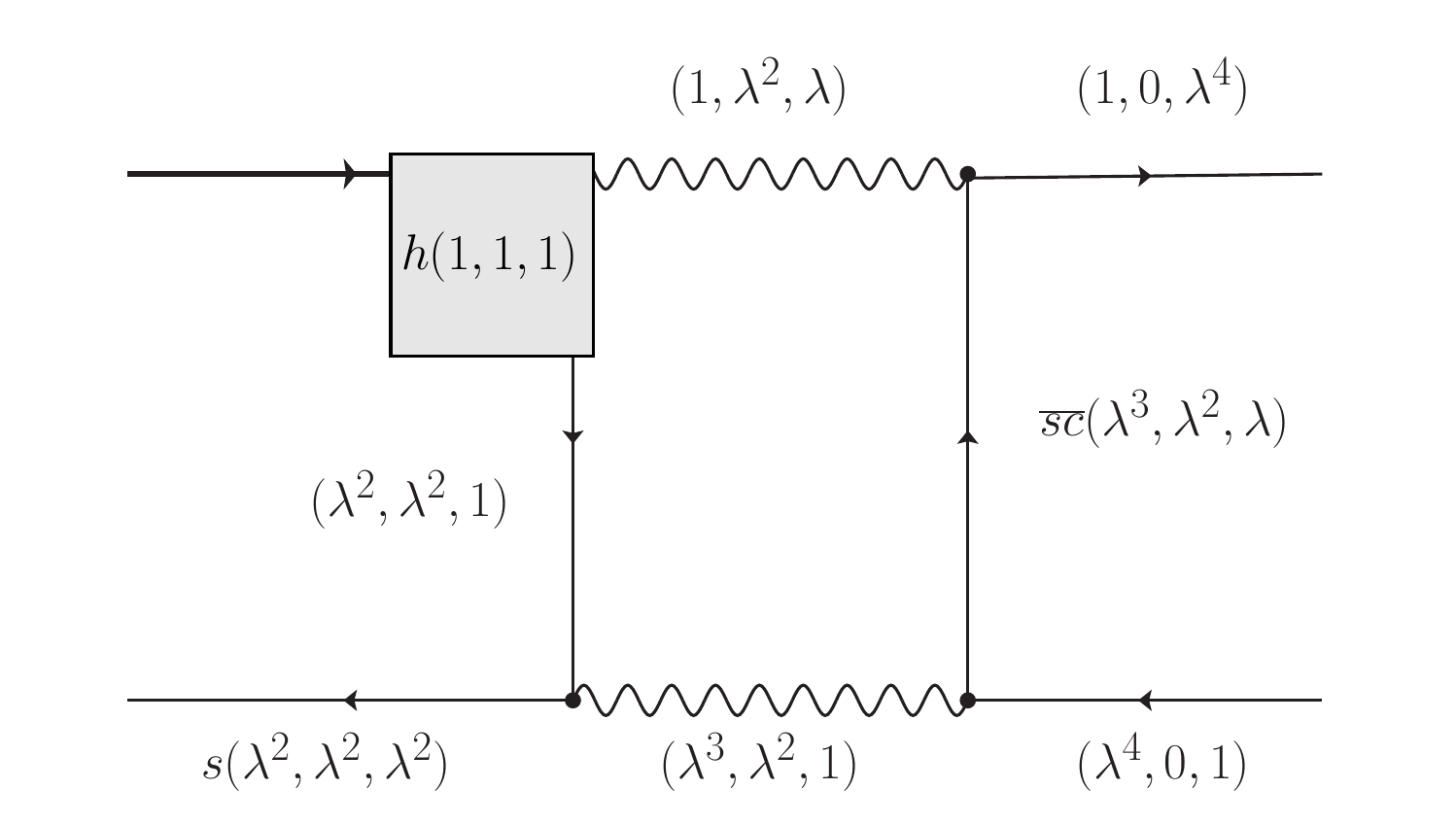}
\hfill 
\includegraphics[width=0.49\textwidth]{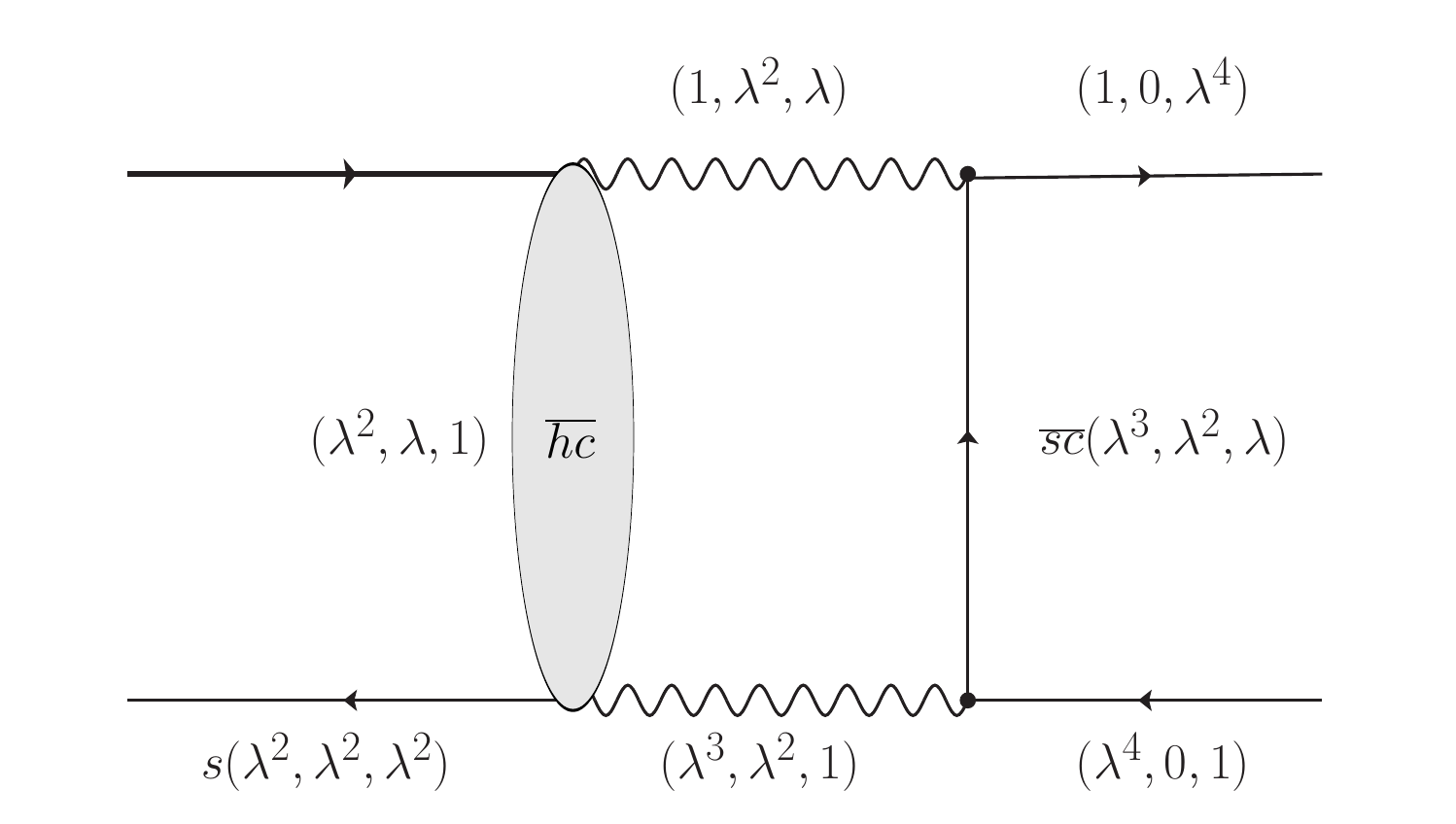}
\end{center}
\caption{\label{fig:QCDcorr} Here we illustrate the QCD corrections to the double-logarithmic enhancement. The left side shows the hard corrections the hard function. The right side shows the correction to the strange-quark propagator.
}
\end{figure}
In this section we concentrate on the fourth line of the renormalized QCD factorization theorem. The aim is to include leading-logarithmic QCD corrections via renormalization group evolution (RGE). In the so-called dual space for the LCDA,
\begin{eqnarray}
	\phi_+(\omega) &=& \int \frac{d\omega'}{\omega'} \, \sqrt{\frac{\omega}{\omega'}} \, J_1\left( 2 \sqrt{\frac{\omega}{\omega'}}\right) \, \rho_+(\omega')
	\, ,
\end{eqnarray}
the RGE will be multiplicative \cite{Bell:2013tfa}. There exist three sources of QCD corrections in the process. Two of them are shown in Fig.~\ref{fig:QCDcorr}. The first are hard corrections to the $b \to s \gamma^*$ vertex (left panel), the second anti-hard collinear corrections to the strange-quark propagator (right panel), and finally soft corrections to the LCDA. These corrections are included in the equation
\begin{align}
	&& {\cal F}(E,m)\Big|_{\rm LL} =
	\frac12 \, e^{V(\mu_{hc},\mu_h)} \, e^{V(\mu_{hc},\mu_0) } 
	\int \frac{d\omega'}{\omega'}\, \ln^2 \frac{m^2}{2E \hat \omega'}  \, \left( \frac{\hat \mu_0}{\omega'} \right)^{-g(\mu_{hc},\mu_0)} \rho_+(\omega',\mu_0) 
	\,, \label{res1} 
\end{align}
which contains the evolution of the functions $H_1(0;\mu)$, ${\cal \bar J}_2(1,\omega';\mu) $ and $\rho_+(\omega';\mu)$ via the RGEs. Here the arbitrary scale $\mu$ drops out as as required by definition. We now define the generating function for the logarithmic moments of the LCDA $\rho_+(\omega';\mu_0)$,
\begin{eqnarray}
    F_{[\rho_+]}(t;\mu_0,\mu_m) &=& 
    \int_0^\infty \frac{d\omega'}{\omega'} 
    \left( \frac{\hat \mu_m}{\omega'} \right)^{-t}  \rho_+(\omega',\mu_0) \,,
\end{eqnarray}
from which we can obtain the necessary double-logarithmic term by taking the second derivative w.r.t.\ $t$. Plugging this into Eq. (\ref{res1}) we find
\begin{eqnarray}
    \!\!\!\!\!\!\!\!\!\!\!\!&& {\cal F}(E,m)\Big|_{\rm LL} \!\!=
    \frac12 e^{V(\mu_{hc},\mu_h)} e^{V(\mu_{hc},\mu_0) } 
    \!\left( \frac{2E\mu_0}{m^2} \right)^{\!-g(\mu_{hc},\mu_0)} \!\!\frac{d^2}{dt^2} F_{[\rho_+]}(t+g(\mu_{hc},\mu_0);\mu_0, \frac{m^2}{2E}) \Bigg|_{t=0} .  \nonumber \\ \label{res2}
\end{eqnarray}
This equation provides a compact master formula for the form factor including leading-logarithmic QCD corrections to the double-logarithmic enhancement from the endpoint configuration of the muon propagator.
\section{Explicit parametrization and numerical estimation}
\begin{figure}[t]
\begin{center}
    \includegraphics[width=0.5\textwidth]{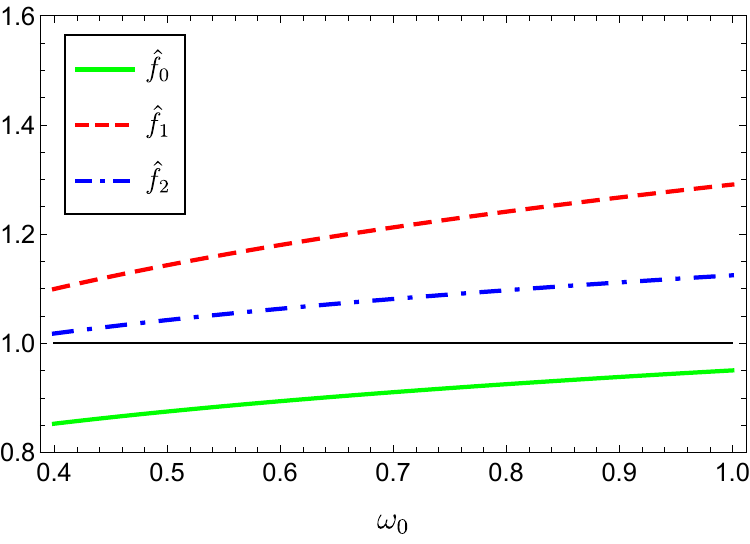}
\end{center}
\caption{\label{fig:num} 
Result of the numerical analysis, that shows the RGE leading-logarithmic QCD effect relativ to the QCD tree-level double-logarithmic enhencment.
}
\end{figure} 
In order to make numerical predictions from the master formula Eq.~(\ref{res2}), we need an explicit but general parametrization of the LCDA as developed recently in  Ref.~\cite{Feldmann:2022uok},
\begin{eqnarray}
    \rho_+(\omega',\mu_0) &=& 
    \frac{e^{-\omega_0/\omega'}}{\omega'} \, 
    \sum_{k=0}^K \frac{(-1)^k \, a_k(\mu_0)}{1+k} \, 
    L_k^{(1)}(2\omega_0/\omega')
    \,,
    \label{eq:generic}
\end{eqnarray}
where the $L_k^{(1)}(2\omega_0/\omega')$ are the associated Laguerre polynomials. The generating function for the logarithmic moments then takes the form,
\begin{eqnarray}
    F_{[\rho_+]}(t;\mu_0,\mu_m) &=& 
    \frac{\Gamma(1-t)}{\omega_0} 
    \left( \frac{\hat \mu_m}{\omega_0} \right)^{-t} \sum_{k=0}^K a_k(\mu_0) \, {}_2F_1(-k,1+t;2;2)
    \,,
\end{eqnarray}
by integrating over $\omega^\prime$. In our analysis we truncate the parametrization at $K = 2$. For this reason we need the first three hypergeometric functions with negative integers $-k$ as first argument,
\begin{equation}
\begin{aligned}
    {}_2F_1(0, 1 + t; 2; 2)
        &= 1 \,, \cr 
    {}_2F_1(-1, 1 + t; 2; 2)
        &= -t \,, \cr
    {}_2F_1(-2, 1 + t; 2; 2)
        &= \frac{1}{3} \left(1 + 2t^2\right) \,,
\end{aligned}
\end{equation}
which have a polynomial structure. Plugging this parametrization into the master formula in Eq.~(\ref{res2}) yields the result for the form factor:
\begin{eqnarray}
	{\cal F}(E,m)\Big|_{\rm LL} &\simeq&
	\frac{\Gamma(1-g)}{2\omega_0} \, e^{V(\mu_{hc},\mu_h)+V(\mu_{hc},\mu_0)}
	\left( \frac{\hat \mu_0}{\omega_0}\right)^{-g} 
	\cr 
	&&  {} \times \left\{ 
	\left(a_0 - g \, a_1 + \frac{1+2g^2}{3} \, a_2 \right) \left[ \left( \ln \hat \lambda_0^2+\psi(1-g)\right)^2 + \psi'(1-g)\right]
	\right.
	\cr && \left. {} + \left( 2 a_1 -\frac83 \, g\, a_2\right) \left( 
	\ln\hat\lambda_0^2  + \psi(1-g) \right) 
	+ \frac{4 a_2}{3} 
	\right\} 
	\equiv \sum\limits_{k=0}^{2} a_k \, f_k(\omega_0)
	\,.
\label{FRGexp}
\end{eqnarray}
Here $\psi$ is the digamma function, 
$a_k=a_k(\mu_0)$,
$g=g(\mu_{hc},\mu_0)$, and we can define 
$$
 \hat \lambda_0^2 \equiv \frac{m^2  e^{2\gamma_E}}{2E\omega_0} \,.
$$
To estimate the effects of the leading-logarithmic QCD corrections, we use $\mu_0=1$~GeV, $\mu_h=5.3$~GeV, $\mu_{hc}=\sqrt{\mu_h \mu_0} \simeq 2.3$~GeV for our numerical analysis. For the strong coupling $\alpha_s$ the values
$$
  \alpha_s(\mu_0)=0.49 \,, 
  \qquad \alpha_s(\mu_h)\simeq 0.21 \,,
$$
at the relevant scales are used, which leads to
$$
    z(\mu_{hc}, \mu_h)=1.30 \,, \qquad z(\mu_{hc}, \mu_0)= 0.65 \,,
$$
for $n_f=4$. This can be used to calculate the RG factors und potentials (see~\cite{Feldmann:2022ixt} for their definition), 
$$
  g \simeq 0.138 \qquad \mbox{and} \qquad V(\mu_{hc}, \mu_h) \simeq -0.037 \,, \quad 
  V(\mu_{hc},\mu_0) \simeq -0.053  \, .
$$
The results of the numerical analysis are shown in Fig.~\ref{fig:num}. We plot the factors $f_k(\omega_0)$ in front of the expansion parameters $a_k$ (compare to Eq. (\ref{FRGexp})) in a normalized form according to
\begin{equation}
    \hat f_k(\omega_0)\equiv \frac{f_k(\omega_0)}{f_k(\omega_0)|_{g=V=0}} \,,
\end{equation}
where for the normalized case we set $g=V=0$. In Fig.~\ref{fig:num} we find the impact of the leading-logarithmic QCD corrections to the corresponding coefficients $a_{0,1,2}$ that show up in the explicit parametrization and in Eq.~(\ref{res2}). This effect can maximally be of the order $(-15,+30,+10)$\%, depending on the scale $\omega_0$. The $\omega_0$-dependence of the RG evolution originates mostly in the factor $(\hat \mu_0/\omega_0)^{-g}$ in Eq.~(\ref{FRGexp}).
A recent estimate of the numerical size of the coefficients $a_{0,1,2}$ for the parametrization of the LCDA for the $B_s$ meson can be found in Ref.~\cite{Feldmann:2023aml}.

\section{Summary and outlook}
We studied the QCD factorization of the ${\cal O}_7$ contribution to the $\bar B_s \to \mu^+\mu^-$ decay amplitude. With the help of the method of regions we calculated the endpoint divergent convolution integrals and regulated them by adding a rapidity regulator. By summing up all the different momentum regions we managed to reproduce the already known result for the decay amplitude. It was possible to identify the specific momentum regions where the double-logarithmic enhancement in a small ratio of soft and hard scales of the process shows up. We were also able to construct the bare factorization theorem and perform the refactorization procedure. Hence we could write down the renormalized QCD factorization theorem for the ${\cal O}_7$ contribution to the decay amplitude. Then concentrating on the part of the factorization theorem that contains the double-logarithmic enhancement we included leading logarithmic QCD corrections using the renormalization group evolution. We were able to provide a useful master formula for these corrections. Using a parametrization for the light cone distribution amplitude we provided a numerical analysis to estimate the leading logarithmic QCD corrections. These can be of ${\cal O}(30 \%)$ relative to the leading order $\alpha_s^0$ contribution. The next steps should be to calculate the hard and anti-hard-collinear functions at ${\cal O}(\alpha_s)$ fixed order. Additionally, it would be interesting to conceptually understand the factorization of QED corrections.

\begin{acknowledgments}
I would like to thank the organizers of FPCP 2023 for the opportunity to present our work. I would also like to thank Thorsten Feldmann, Nico Gubernari and Tobias Huber for comments on the manuscript. This research is supported by the Deutsche Forschungsgemeinschaft (DFG, German Research Foundation) under grant 396021762 --- TRR 257. Diagrams were drawn with JaxoDraw~\cite{Binosi:2003yf}.
\end{acknowledgments}

\end{document}